# Channel-wise Feature Decorrelation for Enhanced Learned Image Compression

Farhad Pakdaman, *Member IEEE*, and Moncef Gabbouj, *Fellow, IEEE*

*Abstract*—The emerging Learned Compression (LC) replaces the traditional codec modules with Deep Neural Networks (DNN), which are trained end-to-end for rate-distortion performance. This approach is considered as the future of image/video compression, and major efforts have been dedicated to improving its compression efficiency. However, most proposed works target compression efficiency by employing more complex DNNS, which contributes to higher computational complexity. Alternatively, this paper proposes to improve compression by fully exploiting the existing DNN capacity. To do so, the latent features are guided to learn a richer and more diverse set of features, which corresponds to better reconstruction. A channel-wise feature decorrelation loss is designed and is integrated into the LC optimization. Three strategies are proposed and evaluated, which optimize (1) the transformation network, (2) the context model, and (3) both networks. Experimental results on two established LC methods show that the proposed method improves the compression with a BD-Rate of up to 8.06%, with no added complexity. The proposed solution can be applied as a plug-and-play solution to optimize any similar LC method.

## I. Introduction

THE emerging Learned Compression (LC) replaces traditional codec modules with a parametric model such as Deep Neural Networks (DNNs), which is trained to learn compression. This is in response to the ever-increasing demand for multimedia content, which motivates innovations for more efficient compression schemes [1]. Although LC is a rather new direction and is still finding its shape, recent LC methods [2][3][12][10][13] have shown great potential and achieved high compression efficiency, overtaking the state-of-the-art traditional codecs such as HEVC [4] and VVC [5].

Pioneering work of Ballé et al. [6][7] and Theis et al. [8] were among the early LC methods, which formulate compression as Rate-Distortion Optimization (RDO), and solve it iteratively. CNN and RNN architectures are used to transform the image into a latent feature representation, which is then compressed with entropy coding, using a context model. To improve the compression, great efforts have been dedicated to designing advanced (and usually more complex) network architectures [9][10], and context models [3][7][11][12]. Ballé et al. [7] proposed an auxiliary hyperprior network that extracts scale parameters to capture spatial dependencies in latent representation. Minnen et al. [11] presented an autoregressive and hierarchical context model, to better capture the dependencies. Cheng et al. [3] extends this by modeling the context with discretized Gaussian mixtures. Several works build on these context models to propose improved compression. Operational neural networks with built-in non-linearity [9], transformer-based architecture [10], and diffusion-based models [13] are among novel advanced architectures. Flexible bitrate LC [14][15], compressing for machine vision [16], latent-space analysis [17], and advances in learned video compression [18][19][20] are among other developments and applications of LC.

A main downside of LC methods is its high computational complexity which can be tens to hundreds of times more than traditional codecs [21]. Most of this complexity comes from using complex DNNs for the codec, and complex context models. To mitigate the complexity, recent proposals suggested network pruning [22][23] to reduce the computations, or channel grouping [24] and checkerboard context modeling [25] for improved parallelization. Although these works reduce the complexity, the reduction is not enough for real-time or energy efficient codecs, and this reduction is compensated by the emerging research proposing more complicated DNNs.

To mitigate this issue, we set to enhance the compression efficiency by better training the existing LC architectures and fully exploiting the DNN capacity. Recent research has shown that learning a divers set of features leads to higher performance [26]. This issue has been studied in the context of learning diverse parameters [27], and more recently diverse features [28][29], with the latter showing more promising results.

Inspired by above mentioned works, we propose an optimization approach for LC, which further exploits the existing DNN capacity without adding to the computational complexity. We argue that the existing feature diversity methods are not efficient for LC, as (1) diversifying information within each feature channel complicates image reconstruction and is not effective, and (2) the large latent size of LC makes it inefficient to decorrelate all features. Hence, we propose to improve feature diversity in codec, by decorrelating feature channels at the encoder bottlenecks. The RDO is modified to include the proposed decorrelation loss as a regularization term. We propose three strategies for the proposed method which (1) optimize the transformation network via decorrelating

Manuscript received xx Februeary 2024; revised xx xxx xxxxx; accepted xx xxx xxxxx. Date of publication xx xxx xxxxx; date of current version xx xxx xxxxx. This project has received funding from the European Union's Horizon 2020 research and innovation programme under the Marie Skłodowska-Curie grant agreement No 101022466.

Authors are with the Faculty of Information Technology and Communication Sciences, Tampere University, 33101 Tampere, Finland (e-mail: farhad.pakdaman@tuni.fi; moncef.gabbouj@tuni.fi).

transformed image features, (2) optimize the context (auxiliary) network by decorrelating context features, and (3) optimize both networks. The proposed optimization method can be applied to any LC which is optimized for Rate-Distortion (R-D) performance. Experimental results confirm that an improved compression performance can be achieved with no added complexity. Contributions of this work are:

- We propose to improve compression by improving feature diversity at the LC's latent representation. This exploits the existing network capacity without adding complexity.
- We design a new loss function and reformulate the Rate-Distortion Optimization to guide the training towards more diverse features.
- We propose three strategies to optimize the transformation network, context network, and both.
- We evaluate the proposed method and provide detailed experimental results on two established LC methods.

## II. Proposed Method

Fig. 1 (a) depicts a generic architecture for LC compression method. The input image $x$ is transformed into a latent representation $y$, which is quantized into $\hat{y}$ and transmitted to the decoder after Arithmetic Encoding (AE). A side bitstream is formed by further processing $y$ with a hyper-encoder, to extract context parameters used for entropy coding. Fig. 1 (b) visualizes an example for latent representations $y$ and $z$. These representations include feature channels which capture different signal patterns and help reconstructing a robust image. However, the usual optimization of LC methods leads to correlated features, which cannot fully benefit from the DNN capacity. In following subsections, we formulate the optimization problem and detail the proposed method.

*A. Learned Image Compression Framework.*

We start formulating the problem using a similar framework as in [3]. Given an input image $x$, a transformation (or analysis) network $g_a(.)$ is used to derive latent representation $y$. This latent is then quantized and transmitted to decoder after entropy coding. At decoder, a synthesis network $g_s(.)$ is used to reconstruct the image, $\hat{x}$. This process is given as following, where $\phi$ and $\theta$ are the learned parameters of the analysis and synthesis networks.

$$y = g_a(x; \phi) \quad (1)$$

$$\hat{y} = Q(y) \quad (2)$$

$$\hat{x} = g_s(\hat{y}; \theta) \quad (3)$$

As quantization is not differentiable, during training it is replaced with uniform noise $U(-0.5, 0.5)$. To capture the dependencies in representation $\hat{y}$, a set of auxiliary analysis and synthesis networks, $h_a$ and $h_s$ are used, which model the feature distribution $p_{\hat{y}|\hat{z}}(\hat{y}|\hat{z})$, to derive the entropy parameters for AE. For $\hat{z}$ a factorized prior is used [7]. This process is given as:

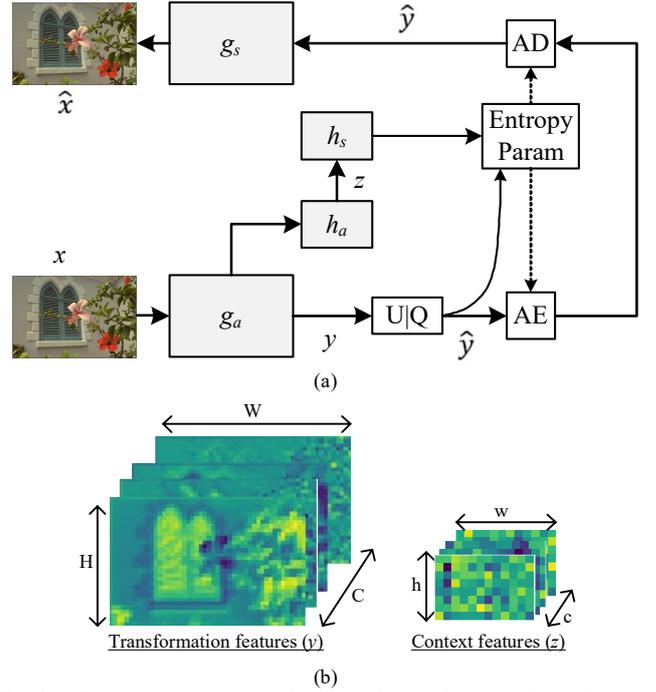

Fig. 1.(a) Learned compression architecture, (b) visualization of latent features

$$z = h_a(y; \phi_h) \quad (4)$$

$$\hat{z} = Q(z) \quad (5)$$

$$p_{\hat{y}|\hat{z}}(\hat{y}|\hat{z}) <= h_s(\hat{z}; \theta_h) \quad (6)$$

To train the encoder and decoder networks, an RDO loss is defined, where the rate, $R$, is approximated by entropy of feature elements, $R(\hat{y}) = \mathbb{E}[-log_2 p_{\hat{y}|\hat{z}}(\hat{y}|\hat{z})]$, while distortion, $D$, is measured as a distance such as Mean Square Error (MSE). This loss is given in (7), where $\lambda_d$ is used to balance between the rate and distortion.

$$\mathcal{L} = R(\hat{y}) + R(\hat{z}) + \lambda_d D(x, \hat{x}) \quad (7)$$

*B. Rate-Distortion Optimization with Feature Decorrelation*

As discussed earlier, the RDO process given by (7) tends to learn correlated features, which cannot fully exploit the network capacity. To mitigate this, we propose to decorrelate the latent features, to guide the encoder to learn a more diverse set of features. Hence, we modify the R-D loss to diversify features for improved performance. The modified RDO loss is given in (8).

$$\mathcal{L} = R(\hat{y}) + R(\hat{z}) + \lambda_d D(x, \hat{x}) + \lambda_{fd} L_{fd} = R(\hat{y}) + R(\hat{z}) + \lambda_d (D(x, \hat{x}) + \alpha L_{fd}) \quad (8)$$

Here, $\lambda_{fd}$ is the weight of feature decorrelation loss. Directly optimizing for the first part of (8) tends to shift the RD curves and complicates the rate allocation. Hence, we reformulate the loss as the second term of (8), such that $\lambda_{fd} = \alpha \lambda_d$, which bounds the effect of the added term to the selected $\lambda_d$ (corresponding to the selected rate). Next subsection details the calculation of the feature decorrelation loss.

*C. Channel-wise Feature Decorrelation*

This section details the feature decorrelation loss. Decorrelating all features as suggested by prior works [28] is not suitable for image compression, as (1) the latent size in compression can be very large, and calculating correlation or covariance matrix for such a large number of elements is often infeasible, (2) our experiments show that decorrelating features within each feature channel hurts the R-D performance and not effective for compression. Hence, we propose to diversify the features along the channels.

Given a latent feature representation, $F \in \mathbb{R}^{N \times W \times H \times C}$, the feature decorrelation loss $C(F)$ for each spatial element is given as (9), where $\bar{F}$ indicates the mean feature value. The loss sums up the pairwise correlation among all channel elements ($C$ channels) in a spatial position in the $W \times H$ plane, over a batch of $N$ image samples.

$$C(F) = \sum_{u,v \in C} |Cov(F^u, F^v)| = \sum_{u,v \in C} |\sum_{i \in N}(F_i^u - \bar{F}^u)(F_i^v - \bar{F}^v)| \quad (9)$$

Accumulating $C(F)$ over all spatial locations, the feature correlation loss $L_{fd}$ is achieved as (10).

$$L_{fd} = \sum_{W \times H} C(F) \quad (10)$$

To apply the proposed feature diversity loss on LC, we propose three strategies, which optimize the transformation networks $g_a/g_s$, hyper (auxiliary) networks $h_a/h_s$, and both. To do so, for the first variant (named **Proposed y**), we apply $L_{fd}$ on $F = y$, and for the second variant (named **Proposed z**), we apply $L_{fd}$ on $F = z$. For the third variant (named **Proposed y+z**), we calculate the sum of loss on both $y$ and $z$; this is $L_{fd} = \sum_{W \times H \ for \ y} C(y) + \sum_{w \times h \ for \ z} C(z)$.

Moreover, the feature diversity loss can be applied either on the feature representations, or on the quantized features. Although both cases were tested successfully, we found empirically that applying directly on features (e.g., on $y$ instead of $\hat{y}$) achieves slightly improved results. This can be justified by the fact that in training, $\hat{y}$ is approximated by added noise and it does not correspond to actual quantized values.

## III. EXPERIMENTAL RESULTS

The proposed method is implemented and tested using CompressAI library [30] in PyTorch. As the proposed method can be used to optimize any LC method, we implemented and tested it on two established LC methods, Cheng's method [3] and the scale hyperprior method [7]. For evaluations, we train the proposed methods and the corresponding baselines ([3][7]) with similar configurations. All methods are trained over patches of 256×256 random samples from the JPEG-AI dataset [31], for 100 epochs and over 30K iterations, and a batch size of 16. MSE loss is used for $D$. The $\lambda_d$ is set to 0.0018, 0.0035, 0.0067, 0.013, 0.025, and 0.0483 to train six models corresponding to different rates. All models use 128 channels.

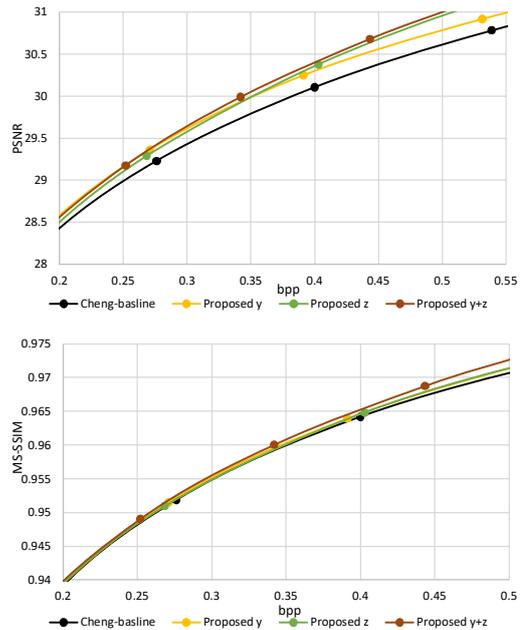

Fig. 2. R-D curves on Kodak, based on PSNR (top) and MS-SSIM (Down)

The $\alpha$ for each model is selected from a range of [$10^{-8}$, $10^{-5}$]. Models are trained on an Nvidia Tesla V100, and tested on a system with Intel Core i7 11850H and Nvidia T1200.

*A. Rate-Distortion Performance*

Fig. 2. Compares the R-D performance of the proposed methods with their corresponding baseline [3], using PSNR and MS-SSIM, on Kodak dataset. It can be observed that all three variants of the proposed method consistently outperform the baseline, leading to bitrate saving in similar qualities. To quantify the performance gain, we measure Bjøntegaard Delta Rate (BD-Rate) [32] of each proposed method against baseline, based on PSNR and MS-SSIM. Table I summarizes the BD-Rates, which are inline with the curves. It can be observed that the Proposed y achieves slightly higher bitrate reduction compared to the Proposed z. Merging the two methods, Proposed y+z outperforms both individual variants, however, the bitrate savings from the two does not accumulate. When measured by MS-SSIM, the gains are lower than the PSNR based measurements. This is justified as the models are trained based on MSE, which is closely correlated to PSNR. The best variant, Proposed y+z, achieves a considerable BD-Rate gains of -8.06% based on PSNR, and -2.74% based on MS-SSIM.

TABLE I. BD-RATES COMPARED WITH THE BASELINE [3]

|  | BD-Rate PSNR (%) | BD-PSNR | BD-Rate MS-SSIM (%) | BD-MSSSIM × 10² |
|---|---|---|---|---|
| Proposed y | -6.94% | 0.178 | -1.62% | 0.065 |
| Proposed z | -6.76% | 0.194 | -1.31% | 0.055 |
| Proposed y+z | -8.06% | 0.206 | -2.74% | 0.116 |

To further evaluate the performance, we also implement the Proposed y method for hyperprior method [7], and compare the results with the hyperprior baseline. Table II shows that the proposed method succeeds in improving the baseline results by 1.17%. The achieved gain is lower compared to models developed based on Cheng's method [3]. This is justified as hyperprior method uses a much simpler network, with only around 0.4 the number of parameters [21] of Cheng, and a simpler context model. Hence, the more sophisticated Cheng has more capacity to benefit from feature decorrelation.

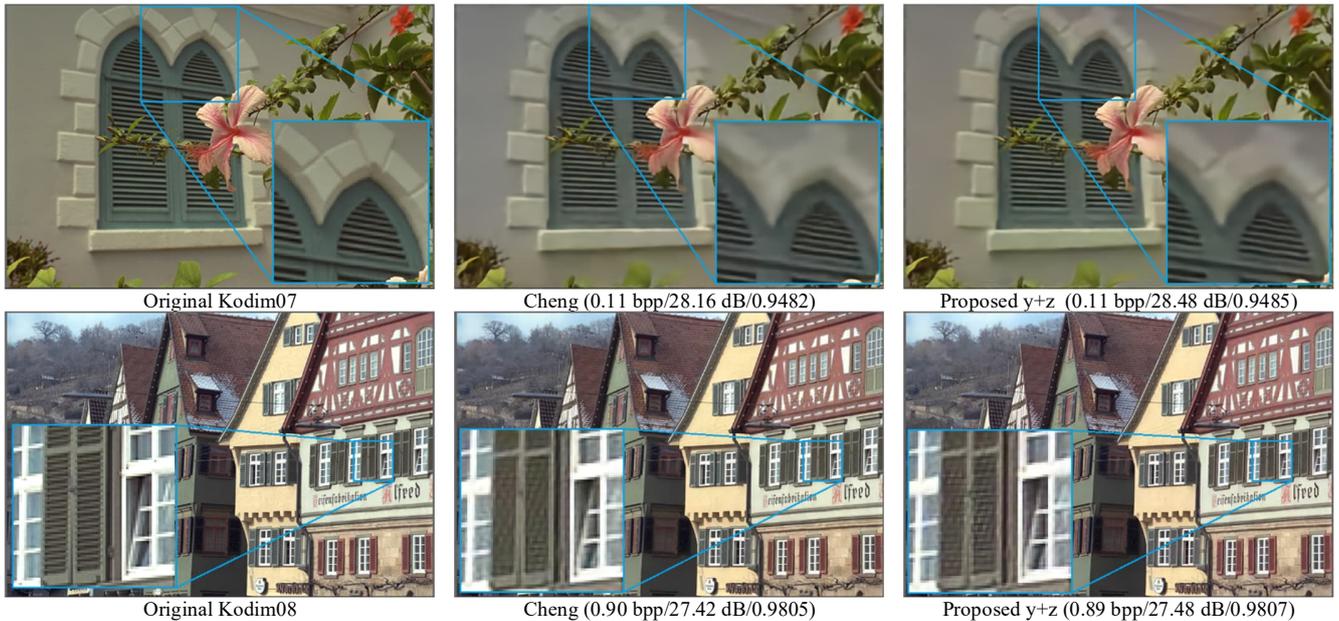

Fig. 3. Comparing visual quality of the proposed y+z method, with its Cheng [3] baseline and the original image. Values are bpp/PSNR/MS-SSIM, respectively.

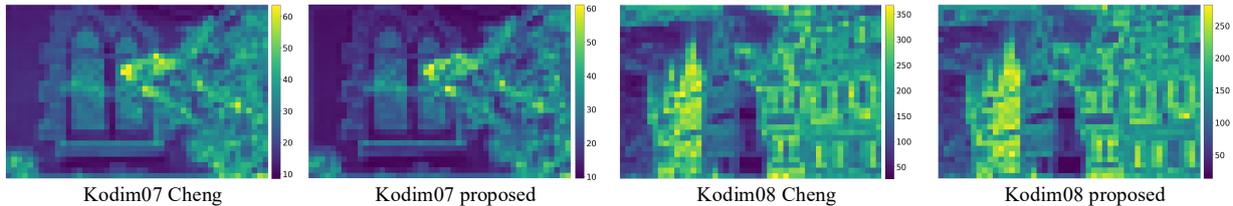

Fig. 4. Feature representations (sum of all y channels) for the baseline and the proposed y+z method

TABLE II BD-Rates compared with the baseline [7]

|  | BD-Rate PSNR (%) | BD-PSNR | BD-Rate MS-SSIM (%) | BD-MSSSIM $\times 10^2$ |
|---|---|---|---|---|
| Proposed y | -1.17% | 0.0411 | -0.71% | 0.021 |

### B. Visual Quality

To further assess the proposed method, we compare its visual quality with the baseline [3], and original images in Fig. 3. Examples were selected to have similar bpps. It can be observed that the proposed y+z has a higher visual quality and preserves more details, for instance in building structures or on window blinds. Analyzing the learned features in Fig. 4 reveals that the feature maps (sum of all y channels) of the two methods show similar structure, while the proposed method exhibits slightly more details in complex areas. However, measuring the correlation among feature channels reveals that the proposed method indeed learns more distinct features, which is the main cause for its higher R-D performance. Sum of correlation among all channels for the proposed method are 0.51 and 0.42 those of Cheng, for examples of kodim07 and kodim08 images.

### C. Computational complexity

As discussed in section I, the proposed method can be applied to any LC method which is optimized with an RDO process, and does not change the network architecture or the context model. Hence, it improves the R-D performance without adding to the encoding/decoding complexity. However, computing an extra term for the feature decorrelation loss slightly increase the training time. As summarized in Table III, the proposed methods take 5.32% to 9.97% longer in training.

TABLE III Training time increase (%) compared to baseline

|  | Proposed y | Proposed z | Proposed y+z |
|---|---|---|---|
| Training time | 6.98% | 5.32% | 9.97% |

### IV. CONCLUSION

This paper proposed an improved method for training learned image compression, which improves the compression performance by learning more diverse features in encoder latent. It was pointed that traditional rate-distortion optimization does not fully exploit the network's capacity, due to high correlation among learned features. To remedy this, a channel-wise feature decorrelation loss was designed that aims to guide the training to diversify feature channels. Moreover, the designed loss was integrated into the Rate-Distortion Optimization, to design three variants. Experimental results confirmed that all variants consistently improve the compression performance, while optimizing both transformation and context networks achieves the best results. The proposed method does not affect the encoder/decoder complexity after training.